\documentclass[11pt]{article}
\usepackage{moriond,epsfig}

\def\Journal#1#2#3#4{{#1} {\bf #2}, #3 (#4)}

\newcommand{\hetrois}    {\mbox{$ ^{3}{\mathrm{He}}                            $}}

\newcommand{\neutt}{$\tilde{\chi}$}

\def\NIMA{{\em Nucl. Instrum. Methods} A}

\def\PLB{{\em Phys. Lett.}  B}

\def\PRD{{\em Phys. Rev.} D}

\def\be{\begin{equation}}
\def\ee{\end{equation}}
\def\bea{\begin{eqnarray}}
\def\eea{\end{eqnarray}}

\begin{document}
\vspace*{4cm}
\title{Dark matter directional detection with MIMAC}

\author{C. Grignon$^1$, J. Billard$^1$, G. Bosson$^1$, O. Bourrion$^1$, O. Guillaudin$^1$, C. Koumeir$^1$, F. Mayet$^1$, D.~ Santos$^1$, P. Colas$^2$, E. Ferrer$^2$, I. Giomataris$^2$ }
\address{$^1$ LPSC, Universit\'e Joseph Fourier Grenoble 1,
  CNRS/IN2P3, Institut Polytechnique de Grenoble, 53 avenue des Martyrs, 38026 Grenoble, France \\        
 $^2$ IRFU/DSM/CEA, CE Saclay, 91191 Gif-sur-Yvette cedex, France}

\maketitle
\abstracts{
MiMac is a project of micro-TPC matrix of gaseous ($^3$He, $\rm CF_4$) chambers for direct detection of non-baryonic dark matter.
Measurement of both track and ionization energy will allow the electron-recoil discrimination, while access to the directionnality of the tracks 
will open a unique way to distinguish a geniune WIMP signal from any background.
First reconstructed tracks of 5.9 keV electrons are presented as a proof of concept.}

\section{Introduction}

Nowadays, there is a strong evidence in favor of a  dark  matter dominated Universe :  locally,  
from the rotation curves of spiral galaxies \cite{rubin} or
the Bullet cluster \cite{clowe} and on the largest scales, from  cosmological  observations \cite{wmap,archeops}. 
Most of the matter in the Universe consists of cold non-baryonic dark matter (CDM), the 
leading candidate for this class of yet undiscovered particles (WIMP)  being the 
lightest supersymmetric particle. 
In various supersymmetric scenarii (SUSY), this neutral and colorless particle is the lightest 
neutralino \neutt.\\
In order to detect this particle, tremendous experimental efforts on a host of techniques have been made.
Whatever the detection technique used, ultimately, the problem is to distinguish a geniune WIMP event from  backgrounds (mainly neutrons 
and $\gamma$-rays). 
The most promising  strategy is to search for a favored incoming direction for the WIMP signal, the sun's velocity vector beeing oriented 
towards the Cygnus constellation \cite{directionality,agreen}. 
Several projects aiming at directional detection  of Dark Matter are being developped \cite{Drift,mit,MIMAC,Collar}.\\
Gaseous $\mu$TPC detectors present the privileged features of being able to reconstruct 
the track of the recoil following the interaction, thus allowing to access both the 
energy and the track properties (lenght and direction). 
Although a precise  measurement of the energy of the recoil is the starting point of any background discrimination, the 3D reconstruction of the track
is necessary to do some dark matter directional detection.

\section{The MIMAC project}

The MIMAC project is a multi-chamber detector for Dark Matter search. The idea is to measure both track and ionization with 
a matrix of micromegas $\mu$TPC filled with \hetrois~and $\rm CF_4$. The use of these two gases is motivated by their privileged features for dark matter search. In  particular, a detector made of such targets will be sensitive to the spin-dependent 
interaction, leading to a natural complementarity with existing detectors mainly sensitive to 
scalar interaction, in various SUSY models, e.g. non-universal SUSY \cite{mayet-susy,moulin}.  
Using both  \hetrois~and $\rm CF_4$ in a patchy matrix of $\mu$TPC opens the possibility to compare rates for two atomic masses, 
and to study neutralino interaction separately with neutron and proton as the main contribution to the spin content of these nuclei. 
With low mass targets, the challenge is to measure low energy recoils, below 6 keV for Helium, by means of ionization measurements, and to reconstruct the 
3 dimentional track of the nuclei during low pressure operation of the MIMAC detector.
Combining these two informations will open the possibility to discriminate neutralino signal and background 
on the basis of track features and directionality.

\begin{figure}[t]
\begin{center}
\includegraphics[scale=0.4]{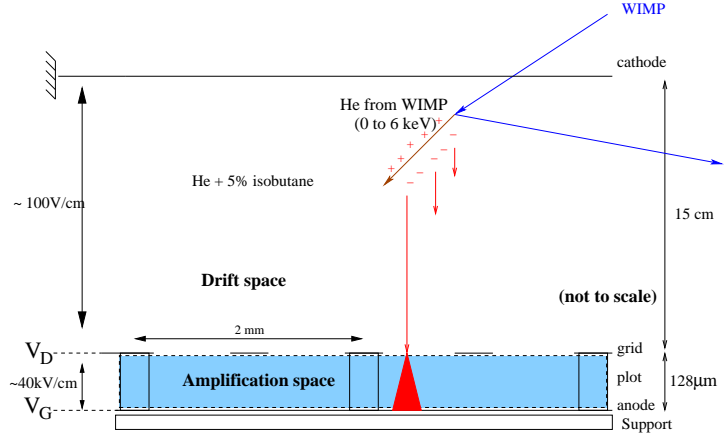}
\caption{Sketch of the Micromegas $\mu$TPC used for the Quenching factor measurement (not to scale).}
\label{fg.micromegas}
\end{center}
\end{figure}

\section{Measurement of the ionisation quenching factor}

In order to distinguish signal from background, it is essential to precisely measure the energy of the recoil.
This energy is released in the detection medium and is shared among three different processes: ionization, scintillation and heat.
The fraction of energy given to electrons is defined as the ionization quenching factor (IQF) and has been estimated theoretically \cite{Lindhard} 
and parametrized by \cite{Lewin}, but has never been measured at low energy in helium gaz, due to 
ionization threshold  of detectors and experimental constraints.\\
As described in \cite{prl},\cite{ogparis},\cite{fmparis}, we designed an Electron Cyclotron Resonance Ion Source, able to produced various 
ions (proton, $^3$He, $^4$He, $^{19}$F) with an energy from 1 keV to 50 keV.
We successfully measured the IQF in helium at low energies, and reached an energy threshold below 1 keV, which is a key point for Dark Matter detectors, since it is needed 
to evaluate the nucleus recoil energy and hence the WIMP kinematics.\\

\section{Track reconstruction in 3D}

\begin{figure}[t]
\begin{center}
\includegraphics[scale=0.65]{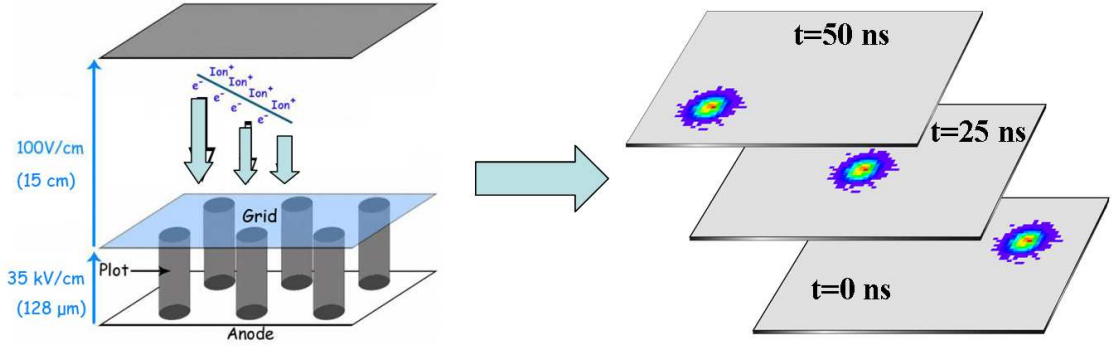}
\caption{Track reconstruction in MIMAC. The anode is scanned every 25 ns and the
3D track is recontructed, knowing the drift velocity,  from the serie of images
of the anode.}
\label{recon}
\end{center}
\end{figure}

The second step of a dark matter project aiming at directional detection is to show the possibility to reconstruct a 3D track. This is a key point as 
the required exposure  is decreased by an order of magnitude between  2D read-out and 3D read-out \cite{agreen}.\\
To perform this 3D reconstruction, we chose to use a segmented anode of a micromegas detector \cite{Giomataris95,bulk} in order 
to collect the electrons produced by the recoil.
As pictured on the figure  \ref{recon}, the electrons move towards the grid in the drift space and are projected on the anode thus allowing to access information on x and y coordinates.
The third coordinate is obtained by sampling the anode every 25 ns and by using the knowledge of the drift velocity of the electrons.

In order to to perform this 40 MHz sampling of each 200 $\mu m$ strip of the anode, we developped a complete electronic system \cite{richer} 
including 16 channels ASICs with a mixer and shaper for energy measurement, a FPGA with on-board processing and DAQ.
We also developped a simulation software to test both the capability of the DAQ to reconstruct tracks and the reconstruction algorithm itself.
For a realistic detector combined to our electronic, we show that the 3D track reconstruction can be achieved with a rather good resolution, 
of $\rm 0.3 mm$ on the track length and below $\rm 4^\circ$ on $\theta$ and $\phi$, assuming a linear trajectory for the recoil ion \cite{simucyril}.\\

\begin{figure}[h!]
\begin{center}
 \includegraphics[scale=0.85]{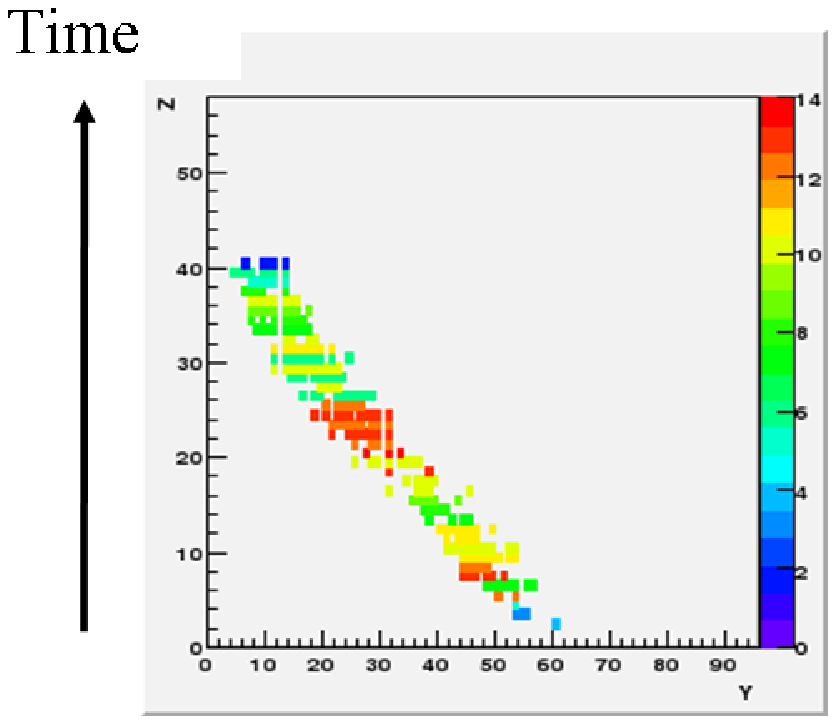}\includegraphics[scale=0.85]{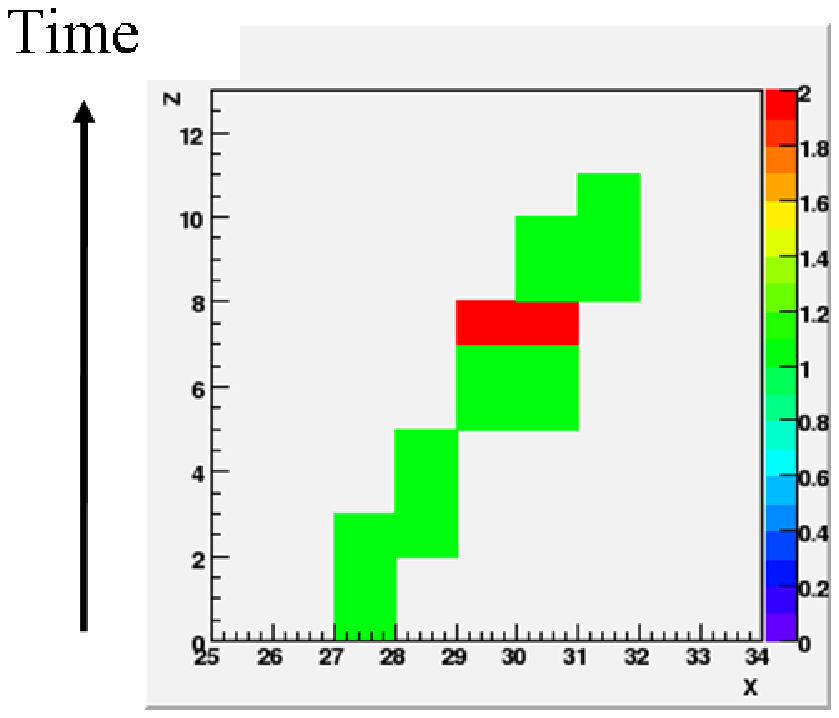}
 \caption{Reconstructed track of a 5.59 MeV $\alpha$ from radon impurities (left) and reconstructed track of a 5.9 keV electron projected on
 the XZ plane (right)}
 \label{electron}
\end{center}
\end{figure}

The first track reconstruction have been obtained with 5.59 MeV $\alpha$ from Radon impurities.  
Figure \ref{electron} presents a projection on a xz plane of an $\alpha$ 3D track obtained at $\rm 800 \ mbar$ in a Helium~+ 5~\% isobutane mixture. 
This can be taken as a proof of the principle of track reconstruction strategy chosen for this project.
Futhermore, we were able to use a sealed alpha source to measure the drift velocity of electrons at a specific E/P value \cite{koumeir}. 

We were also able to reconstruct, for the first time, electron tracks with an energy of 5.9 keV (see figure \ref{electron}), at $\rm 600 \ mbar$.
This highlight the possibility to separate electron event from nuclear recoil, even at very low energy.

\section{Conclusion}

Directional detection offers a robust signature for WIMP detection as long as we have access to the energy and recoil track of the nuclei.
The MIMAC experiment has shown a precise measurement of the recoil energy of ions down to 1 keV 
as well as the ionisation quenching factor of He at this energy range.

Furthermore, we developped a complete electronic system and a dedicated algorithm in order to reconstruct precisely the recoil track in
 3 dimensions.
We were able to reconstruct alpha tracks in the prototype, due to $^{222}Rn$, which can be used to prove the capablity of the detector to
 reconstruct tracks.
Finally, we reconstructed for the first time a 5.9 keV electron track in our prototype, which is a key point for background discrimination in a dark matter
 detection experiment.

The test of the detector will be performed this year with at the Amande facility with a neutron beam of a few keV, in order to check the capability of the 
MIMAC project to reconstruct low energy tracks.

\end{document}